\documentclass[twocolumn,prl,aps,epsfig]{revtex4}

\def\bopi{\mbox{\boldmath $\bar\pi$}}
\def\PI{\mbox{\boldmath $\pi$}}
\def\tvpi{\mbox{\boldmath $\tilde\pi$}}
\def\tpi{\mbox{\boldmath $\tilde\pi$}}
\def\btau{\mbox{\boldmath $\tau$}}

\newcommand{\be}{\begin{equation}}
\newcommand{\ee}{\end{equation}}
\newcommand{\ba}{\begin{array}}
\newcommand{\ea}{\end{array}}

\newcommand{\tC}{{\tilde{C}}}
\newcommand{\tv}{{\tilde{v}}}
\newcommand{\cH}{{\cal H}}
\newcommand{\VM}{{V_{\mu}}}

\newcommand{\bx}{{\bf x}}

\newcommand{\bpi}{{\bar{\pi}}}
\newcommand{\bsig}{{\bar{\sigma}}}
\newcommand{\tsig}{{\tilde{\sigma}}}

\begin{document}
                 

\pagestyle{plain}

\title{
 Linear Sigma 
Model at finite baryonic density and symmetry breakings
}
\author{  {\bf F\'abio L. Braghin} \thanks{e-mail: 
braghin@if.usp.br} \\
{\normalsize Instituto de F\'\i sica, Universidade de 
S\~ao Paulo; C.P. 66.318; CEP 05315-970;  S\~ao Paulo, Brazil.
} } 

\begin{abstract} 
The 
linear sigma model  at finite baryonic density 
 with a massive vector field 
is investigated considering that
all the bosonic fields develop
non zero expected classical values, eventually
associated to condensates and
corresponding to dynamical symmetry breakings
which might occur in  
the QCD phase diagram.
A modified equation for the classical 
vector field is proposed with its respective
 solution.
Some in medium properties of the model (mainly masses) 
are investigated within reasonable prescriptions.
In particular the behavior of the {\it in medium} 
pion and sigma masses and a particular way of 
calculating {\it in medium} coupling
to baryons is investigated.
A symmetry radius for finite baryonic densities
 is proposed and calculated in different ways
in terms of the other variables of the model and 
these different ways of calculating it agree 
quite well.
However, assuming that the pion and sigma masses go to 
zero close to the restoration of chiral symmetry
a too high value for the critical density 
is obtained $\rho_c \simeq
4.3 \rho_0$. 
\end{abstract}

\maketitle

\section{ Introduction}

Matter at high energy densities
has been continuously investigated to provide
a deepeer understanding of strong interactions. 
Experimental
(relativistic and high energy) heavy ion
collisions (r.h.i.c. and h.e.h.i.c.) 
provide very important
collection of data to construct this knowledge.
Asymptotic freedom is a key property 
in this program.
With different approaches, 
hadronic models based on quantum chromodynamics (QCD) 
at finite energy density 
have been extensively investigated and different
condensates are usually expected to appear 
at different  energy densities. 
In the vacuum, the lightest strong interacting particles 
are known to respect, approximatedly at least, chiral symmetry 
$SU_L(2) \times SU_R(2)$ which is expected to be
spontaneously broken down to $SU(2)$.
This is expected to give rise to 
a scalar (quark-antiquark) condensate, 
which might have relevant effects  in 
experimental conditions and in realistic calculations,
it rearranges the theory 
\cite{SSB,TESEDO,VARIATIONAL}. 
As a consequence, the hadronic properties which 
depend on this condensate 
(which can be the order parameter or proportionally
to it \cite{stern}) are 
expected to vary with energy density. 
In particular the behavior of 
the rho vector meson  has been 
considered as a possible signature of the chiral symmetry restoration
via dilepton emission 
\cite{inmedium} although recent experimental 
analysis of
 the rho vector meson spectral 
function shows to be incompatible with energy shift
\cite{NA60-2006}.

In this work 
the Linear Sigma Model (LSM) \cite{LSM} 
at finite baryonic density, 
$\rho_B$, is  investigated with a massive classical
vector field. It is based in \cite{SSBI,IWARA}.
All the mesons  in the model are considered to develop 
classical counterparts.
There is a renewed interest in the pseudoscalar
condensation indeed \cite{SSBI,TOKI-etal,CONDEPI-antigo}.
It provides a way of enhancing CP violation at 
finite density \cite{CP}.
The exact field equations and the stability equation
are truncated such as 
analytical solutions are obtained 
by considering  particular prescriptions for
the stability condition. 
The numerical solutions have a self consistency 
although
the so-called "full self consistency" is only 
achieved in a level of approximation 
and for the interactions considered in a model, not
in the complete self consistency of the exact realistic quantum 
theoretical many body problem which still is too difficult to
obtain.
The truncations in the effective action, done in the
next section, are based in the following considerations:
(1) the effective potential of spin zero bosons keeps  
the same form of that at the tree level calculation 
(i.e., quantum
fluctuations basically rearrange the tree level model),
(2)
each component of the system, i.e. baryons/ spin zero bosons/
spin one fields, have nearly independent stability conditions.
Hopefully this assumption might go along with
the observation of different slope parameters and temperature
freeze-out  \cite{hydro-rhic}
for each of 
the  hadrons emerging from relativistic
heavy ion collisions 
- and eventually different contributions for the
corresponding Hydrodynamics.
The corresponding (dynamical) 
equation for each of the fields is satisfied.
The complete 
numerical investigation of the results will be presented elsewhere
\cite{SSBI,SSBVAC}. 
Several properties of in medium hadrons are 
investigated, namely scalar and pseudoscalar
meson masses and couplings, and their relation to
the behavior of the (chiral) symmetry is 
worked out within a particular prescription which provides
results in agreement with the usually expected behavior.
A symmetry radius is defined for the investigation
of the symmetry behavior and its
estimation (and dependence with the baryonic
density) is done in several ways which yield 
very close result.
In spite of being a quite simple model, without
several degrees of freedom which should be 
relevant at high energy densities, results show
to be consistent.
Assuming that pion and sigma masses go to zero 
close to the density in which chiral symmetry 
would be restored a too high critical density is 
obtained. This can either signal that the model 
is too simple for describing Physics at too high 
densities or that their masses should not be 
expected to be zero close to the phase transition.

\section{The linear sigma model at finite $\rho_B$}

The  Lagrangian density of linear sigma sodel (LSM) with
baryons, $N_i(\bx)$,
sigma and pions, $(\sigma,\PI)$, covariantly
coupled to a  vector field, $\VM$, 
is given by \cite{LSM}:
\be \ba{ll} \label{LSM}
\displaystyle{ {\cal L} = \bar{N}_i (\bx) \left( i \gamma_{\mu}
\tilde{\cal D}^{\mu}  - g_S (\sigma + i \gamma_5 {\btau} . {\PI}  )
\right) N_i (\bx) }\\
\displaystyle{ + \frac{1}{2} \left( {\cal D}_{\mu} \sigma .
{\cal D}^{\mu} \sigma  + {\cal D}_{\mu} {\PI} . {\cal D}^{\mu} {\PI}  
\right) + c \sigma + \frac{1}{2} m^2_V V_{\mu} V^{\mu} +
} \\ 
\displaystyle{ 
- \frac{1}{4} F_{\mu \nu} F^{\mu \nu}  
- \frac{\lambda}{4} \left( \sigma^2 + {\PI}^2 - v^2 \right)^2
,}
\ea
\ee
where the covariant derivatives are given in \cite{LSM}
and they will not be completely considered in this communication. 
The other terms and parameters are standard \cite{LSM,SSBI}.
The introduction of a chemical potential, with
an extra term${\delta \cal L} = - \bar{N} \gamma_0 \mu_{chem} N$,
is nearly  equivalent to 
a shift of the classical temporal component of the 
vector field $V_0$ coupled to the nucleons.
This field however
is a dynamical degree of freedom (d.o.f.) and will be
treated as such.
Since the condensates (such as $\bsig \equiv <\sigma>$)
 depend on the density, so do
most of  the hadronic masses.
Part of the baryon masses are considered to come from the 
the coupling to the scalar mesonic field 
and part from an explicit mass term for the baryons 
in the Lagrangian: $M^* = M \pm g_S \bsig$.

The spin zero fields will be treated in the 
framework of 
the variational Gaussian approach with 
a truncation
\cite{VARIATIONAL,SSBI}.
With the truncation of the effective potential of sigma and pion 
it can be written keeping the 
same form of the tree level effective potential.
The equations for expected values of the sigma and pion 
are found accordingly, and shifts
in these classical parts are considered 
due to the rearrangement 
brought by quantum fluctuations, such as
 $\bsig \to \tsig$.

The total energy density 
is written
in terms  of the four variational parameters for 
the fields $\sigma,\bopi$, 
 plus baryonic densities and vector field variables \cite{SSBI}.
To investigate the behavior the temporal component
of (classical) vector field,
the total energy density is varied with respect to
$V_0$, which is not quantized, instead of 
using its Euler-Lagrange
equation.
In this approach, $V_0$ is found either by writing a reasonable (or exact)
 expression for $\rho_B$ as a function of $V_0$, 
as it is given below, or $V_0$  is treated like
a variational parameter such that one can 
determine a parametric function $\rho_B = \rho_B [V_0]$
which satisfy the equations of mouvement and  stability. 
For this derivation, $\tilde{m}_V$ was kept constant. 
This second procedure 
yields a sort of variational equation for the corresponding
parametric dependence.
The corresponding (variational) equation for a constant
background field component $V_0$ can be given by:
\be \ba{ll} \label{EQmeson}
\displaystyle{ \frac{\partial \cH}{\partial V_0}  = 0 
\;\; \to \;\; g_V \left( \rho_B + 
V_0 \frac{\partial \rho_B}{\partial V_0} \right) - \tilde{m}_V^2 V_0 
= 0.}
\ea
\ee
where the Euler-Lagrange equation for $V_0$
can be recovered by
neglecting the derivative term above.

The stability condition for the ground state, with
binding energy $E_0/A = \cH/\rho_B < 0$, 
can be written as
$
\frac{\partial {\cal H}}{\partial \rho_B} 
  = \left.\frac{ \cal H}{\rho_B} \right|_{\rho_B = \rho_0} < 0,
\; \mbox{and} \;\left.
\frac{\partial^2 \frac{{\cal H}}{\rho_B}}{\partial \rho^2}  
  \right|_{\rho_B = \rho_0} > 0,
$
where $\rho_0$ is the stability density.
The expressions for the energy density and its derivative
with respect to $\rho_B$
 is separated
into three parts  such that each component of the hadronic matter
satisfies the stability equation above
separatedly. With this prescription 
the solutions for the variational equations of each of the 
components satisfy the respective stability equation.
The  reliability of this factorization
is not mathematically proven although 
some arguments for being reasonable
are given  in the
Introduction and in \cite{SSBI}.
The resulting equations (prescriptions) are the following:
\be \label{PRESCR} \ba{ll}
\displaystyle{ (i) \;\;\; 
\frac{\partial E_f}{\partial \rho_B} = \frac{E_f}{\rho_B}; 
\;\;\;\;\;\;  (ii) \;\;\; 
\frac{\partial {\cH}_V}{\partial \rho_B} = \frac{{\cH}_V}{ \rho_B}
 ;
 }\\
\displaystyle{ (iii) \;\;\; 
\frac{\partial (\tsig^2 + \tpi^2 -v^2)}{\partial \rho_B} = 
\frac{(\tsig^2 + \tpi^2 -v^2)}{2 \rho_B} ,}
\ea
\ee
Where ${\cal H}_V = g_V V_0 \rho_B - \frac{1}{2} 
\tilde{m}_V^2 V_0^2 .$
The complete set of 
solutions for the equations will be investigated elsewhere
and compared to the exact numerical solutions. 
The variation of the sigma and pion masses arise from the corresponding 
classical fields. In the vacuum
$\mu_{\pi}^2$ can go to zero as long as $\tpi^2 \to 0$
satisfying the Goldstone theorem when $c \to 0$ in the Lagrangian.

\subsection{ Densities, coupling constants and masses}

The baryon fields, which depend on the bosonic fields through 
the  Dirac equation coupled to the mesons, are quantized in terms 
of creation and annihilation operators.
The baryonic degrees of freedom sum up into 
the densities: baryonic ($\rho_B$),
scalar ($\rho_S$) and pseudo-scalar  ($\rho_{ps}$) densities.
These quantitites will not be explicitely evaluated here
although they are partially used below \cite{IWARA,SSBI}.
The  energy density due to the fermions (antifermions) 
($E_{f,\bar{f}}$)
and the density of baryons (antibaryons)
($\rho_{B,\bar{B}}$)
 can be 
written, in the leading order, in terms of (1) their momenta for 
each kind of baryons ($i$),
up to the last occupied level with momentum $k_F$, and of 
(2) the 
classical vector field as \cite{IWARA,ISMD06}:
\be \ba{ll} \label{RHOF}
E_{(f,\bar{f})}^i  \simeq \frac{\gamma}{ (2 \pi)^3}
 \int^{k_F^i} d^3 k 
\left( \frac{ 2 E_{(+,-)}^i (M^*_i + E_{(+,-)}^i) + 
V_0(V_0 - 2 E_{(+,-)}^i) }{2  
( M^*_i + E_{(+,-)}^i)}
\right) 
\\
\rho_{B,\bar{B}}^i  \simeq \frac{\gamma}{ (2 \pi)^3}
 \int^{k_F^i} d^3 k 
\left( \frac{ (M^*_i + E_{(+,-)}^i)^2 - {\bf k}^2 }{2 M^*_i 
( M^*_i + E_{(+,-)}^i)}
\right)
\
\ea
\ee
In these expressions 
$E_{\pm} = g V_0 \pm \sqrt{k^2 + (M^*)^2}$ 
are the eigenvalues of the 
corresponding 
Dirac equation \cite{IWARA,ISMD06}.
These expressions still correspond to an approximation
and show deviations from the Fermi liquid picture.

The scalar and pseudoscalar densities, which
appear in the equations of $\sigma, \bopi$, can be
expanded in terms of the scalar and pseudoscalar condensates,
for example, as:
\be \ba{ll} \label{rho-expand}
\displaystyle{ \rho_S = \rho_S^{(0)} 
+ \frac{\tsig}{ \tsig_{vac}} \rho_S^{(1)} 
+ \frac{\tsig^2}{ \tsig_{vac}^2} \rho_S^{(2)} + 
\frac{\tsig^3}{ \tsig_{vac}^3} \rho_S^{(3)} + o(|\tvpi|)
,}\\
\displaystyle{ 
\rho_{PS} =  |\tvpi| \rho_{PS}^{(1)} 
+  |\tvpi^2| \rho_{PS}^{(2)} + |\tvpi^3| \rho_{PS}^{(3)} 
+ o(\tsig)
,} 
\ea
\ee
where the coefficients $\rho_{S, PS}^{(j)}$ are
obtained from the expressions calculated with
the solution of the corresponding Dirac equation.
They are such that
$\rho_S = \rho_{ps} = 0$ when $\tsig = \tsig_{vac}$ and $\tvpi=0$,
with higher
order terms are indicated by $o(|\tvpi|)$ and $o(\tsig)$.
By substituting these expressions into the 
variational equations for the respective condensates
 \cite{SSBI}
 it is found that
the terms proportional to $\tsig$ and $\tvpi$ 
can yield contributions
to the {\it in medium} masses of sigma and pions.
The terms proportional to $\tsig^3$ and $\tpi^3$ can produce
(effective) contributions for the coupling constant, i.e.,
$\lambda \to \lambda^* =  \lambda \pm 
\frac{\rho_S^{(3)}}{\tsig^3_{vac}} \simeq
\lambda \pm \rho_{PS}^{(3)}.$
These two corrections for the in medium  effective
coupling constant may also be different from each other,
eventually leading to different interactions of the 
pion and sigma in the baryonic medium.

The second equation of prescriptions (\ref{PRESCR})
can produce the same solution of the
 equation (\ref{EQmeson}).
The third of prescriptions (\ref{PRESCR})  can
define a symmetry radius in 
the medium:
\be \label{RAIQUI} \ba{ll}
\displaystyle{ (\tsig^2 + \tpi^2 -v^2 ) = \tC \sqrt{\rho_B}
.}
\ea
\ee
In this expression $\tC$ is  a constant to be determined from 
the parameters of the model.
In the vacuum:  $\tsig^2 = v^2 = f_{\pi}^2$ as discussed
above.
Modifications in the equation (\ref{PRESCR}-$(iii)$) 
will produce different
dependences on the baryonic density. 
A more general symmetry radius,
corresponding to particular  modifications 
of the corresponding differential equation due to diverse
couplings for example,
might be written as: $(\tsig^2 + \tpi^2 -v^2 ) = 
(\tilde{D} + \tC {\rho_B}^{c})^{\gamma}$ where
$\tilde{D}, \tilde{C}, c, \gamma$ are constants to be related to
the parameters of the model.

One way of calculating $\tC$ is found 
by assuming that this symmetry radius is
valid over a  range of baryonic densities.
This is a crude approximation because, 
heavier hadrons as well as quark and gluon d.o.f. are expected
to be relevant for high energy densities.
In the high density when chiral symmetry should
be restored: $\tsig = \tpi \to 0$.
This critical density is written as
$\rho_c = u \rho_0,$.
Thus in this point:
$\tC = \mp \tv^2 \sqrt{\frac{1}{u .\rho_0}}. $
At the saturation density 
($\rho_B$) the expression for $\tilde{C}$
can be written as:
\be \ba{ll} \label{tsig-tC}
\displaystyle{ \tsig^2 + \tvpi^2 \simeq \tv^2 \left( 1 \pm
\sqrt{\frac{1}{u}} \right) = 
(f_{\pi}^0)^2 \left( 1 \pm 
\sqrt{\frac{\rho_B}{\rho_c}} \right).
}
\ea
\ee
Four values are  considered: 
(i) $u = 2$, (ii) $u= 3$, (iii) 
$u= 3.5$ and (iv) $u=4$.
Considering the branch of solutions for which 
$\tsig^2 + \tvpi^2 < \tv^2$, at $\rho_B = \rho_0$, 
it follows respectively:
\be \ba{ll} \label{estimat}
\displaystyle{ 
\left. \sqrt{\tsig^2 + \tpi^2}
\right|_{\rho_0} \simeq 0.54  \tv  \;\; (i), \;\;\;\;\;\;\;
\left. \sqrt{\tsig^2 + \tpi^2} \right|_{\rho_0} 
\simeq 0.65 \tv \;\; (ii),} \\
\displaystyle{
\left. \sqrt{\tsig^2 + \tpi^2} \right|_{\rho_0} 
 \simeq 0.68 \tv \;\; (iii),
\;\;\;\;\;\;\;
\left. \sqrt{\tsig^2 + \tpi^2} \right|_{\rho_0} 
 \simeq 0.71 \tv \;\; (iv)
.
}
\ea
\ee
Other  solutions are not presented.
Since  the squared value $\bpi^2$ is a scalar which appear very
often in the expressions,
it may be that $f_{\pi}^* \simeq  \sqrt{\tsig^2 + \tpi^2}$,
i.e., a pion classical field could be responsible for 
modifications in the pion decay constant and consequently
measurable, even if competing with other effects.
The topological Skyrme model can provide some argument
in favor of such interpretation 
for a classical pion field
inside hadrons \cite{SKYRMION}.
These expressions may
be therefore useful for relating descriptions of
different ranges of the matter phase diagram.

The values obtained for $\tC$ from estimates (\ref{estimat})
are respectively given by: 
\be \ba{ll} \label{tC4}
\displaystyle{ \tC \simeq \pm 0.41 \mbox{fm}^{-\frac{1}{2}} \;\;(i), 
\;\;\;\;\;\;\; 
\tC \simeq \pm 0.33 \mbox{fm}^{-\frac{1}{2}}  \;\;(ii),} \\
\displaystyle{ 
\tC \simeq \pm 0.30 \mbox{fm}^{-\frac{1}{2}}  \;\;(iii), \;\;\;\;\;\;
\tC \simeq \pm 0.28 \mbox{fm}^{-\frac{1}{2}}  \;\;(iv).
}
\ea
\ee

Another way of estimating $\tC$ is shown by
considering 
the meson masses in the medium. 
With the expressions for meson masses in 
terms of the classical fields and $v$
\cite{SSBI,LSM}, the symmetry radius 
 can be written as:
$\tC \sqrt{\rho_B}  =
\frac{1}{4 \lambda}
( (\mu_T^*)^2 - (\mu_T^{vac})^2 ) 
,$
where $(\mu_T^{(*)})^2 = (\mu_S^{(*)})^2 + 
(\mu_P^{(*)})^2$ at a given density
$\rho_B$.  
In these expressions the coupling $\lambda$ was also
kept constant (and positive) and $c=0$, in the Lagrangian
term.
Two possible behaviors are obtained in this 
picture for the restoration of chiral symmetry: 
the sum of these masses may decrease or increase
depending on the sign of $\tC \sqrt{\rho_B}$.
For  $\rho_0 = 0.15$fm$^{-3}$ and 
$\tC \simeq - 0.15$fm$^{-\frac{1}{2}}$ 
the above expression yields approximated values
$(\mu_T^*)^2 (\rho_0) \simeq  ( 1 \pm 0.53) \;
 \mu_T^{2} (\rho_B=0)$.
If one considers that the pion and sigma masses 
disappear close to the chiral symmetry 
restoration point (i.e., if $(\mu^*)^2 \to 0$), 
with the values above we obtain that 
$\rho_c \simeq 4.3 \rho_0$.
Seemingly it is a too high baryonic density and the reasons
are quite apparent. Firstly, as emphasized above,
the present work only takes into account the light sector
of hadrons and it does not consider quark and gluon 
degrees of freedom. Furthermore, it 
is a controversial subject
 whether pion and sigma masses (two point 
Green's functions) should
be expected to be so close to zero (as it was
assumed to obtain such high value for 
the critical density) close
to (and at the) deconfinement critical point.

These ways of calculating $\tC$
 provide crude (but interesting and curious) estimations.
The corresponding {\it in medium} hadron properties 
are qualitatively  in agreement with other estimations
\cite{inmedium}.
The inclusion of other relevant
d.o.f. will be presented elsewhere as well as 
a corresponding calculation at finite temperature.

\subsection{Summary and Conclusions}

In this work some aspects of 
the Linear Sigma Model 
at finite baryonic density
were investigated with a massive classical
vector field, based in \cite{SSBI,IWARA}.
All the mesons  in the model were considered to develop 
classical counterparts.
In part this is due to 
independent new investigations on pseudoscalar condensates which have shown a renewed interest in the pseudoscalar
condensation indeed \cite{SSBI,TOKI-etal,CONDEPI-antigo}.
The exact field equations and the stability equation
were truncated for obtaining 
analytical solutions which capture the 
expected behavior of the system.
These solutions have a self consistency 
although
the so-called "full self consistency" is only 
achieved in a level of approximation 
and for the interactions considered in a model, not
in the complete self consistency of the exact realistic quantum 
theoretical many body problem which still is too difficult to
obtain.
The truncations in the effective action, done in the
next section, are based in the following considerations:
(1) the effective potential of spin zero bosons keeps  
the same form of that at the tree level calculation 
(i.e., quantum
fluctuations basically rearrange the tree level model),
(2)
each component of the system, i.e. baryons/ spin zero bosons/
spin one fields, have nearly independent stability conditions.
Hopefully this assumption might go along with
the observation of different slope parameters and temperature
freeze-out for each of 
the  hadrons emerging from relativistic
heavy ion collisions 
- and eventually different contributions for the
corresponding hydrodynamic.
The corresponding (dynamical) 
equation for each of the fields are satisfied.
The complete 
numerical investigation of the results will be presented elsewhere
\cite{SSBI,SSBVAC}. 
Several properties of in medium hadrons were 
investigated, namely scalar and pseudoscalar
meson masses and couplings, and their relation to
the behavior of the (chiral) symmetry is 
worked out within a particular prescription which provides
results in agreement with the expected behavior.
A symmetry radius was defined for the investigation
of the symmetry properties and its
estimation (and dependence with the baryonic
density) is done in several ways with fair agreement,
in spite of being a quite simple model, without
several degrees of freedom which should be 
relevant at high energy densities.
Related aspects to 
 matter-antimatter asymmetry in 
relativistic heavy ion collisions and 
in the Early Universe
will
be discussed and investigated  elsewhere \cite{ISMD06}.


\noindent {\Large {\bf Acknowledgements}}

This work was partially supported by FAPESP, Brazil. 
F.L.B. thanks  brief discussions with L.McLerran, R. Rapp, D. Zschiesche,
H. Stoecker, R. Pisarski, M. Munhoz  and 
 G. Krein.

\end{document}